\documentstyle[12pt,psfig,aaspp4]{article}
\newcommand{\beq}{\begin{equation}}
\newcommand{\eeq}{\end{equation}}

\def\_#1{_{\scriptscriptstyle #1}}
\def\^#1{^{\scriptscriptstyle #1}}
\def\pd#1#2{{\partial#1\over\partial#2}}

\def\({[}
\def\){]}
\def\i{\nu}
\def\rhos{\rho^*}
\def\Sigmas{\Sigma^*}
\def\grm{{\rm g}}
\def\gnrm{{\rm g\_{N}}}

\def\div{\vec\nabla\cdot}
\def\grad{\vec\nabla}

\def\a0{a_{o}}

\def\f{\varphi}

\def\abs#1{\vert #1\vert}
\def\ags{\abs{\grad\psi}}

\def\Skr{\Sigma\_{K}(r)}
\def\Sd{\Sigma\_{D}}

\def\d{\delta}
\def\z{\zeta}
\def\vh{v\_{h}}
\def\rhsq{(r\^{2}+h\^{2})}

\def\vg{{\bf g}}
\def\vh{{\bf h}}

\def\vR{{\bf R}}
\begin{document}

\title{The shape of ``dark matter'' halos in disc galaxies according to the 
modified dynamics}
\author{ Mordehai Milgrom}
\affil{Department of Condensed Matter Physics, Weizmann Institute of
Science, Rehovot 76100 Israel}

\begin{abstract}
Analyses of halo shapes for disc galaxies are said to give incongruous
results. I point out that the modified dynamics
(MOND) predicts for disc galaxies a distribution of fictitious dark matter
that comprises two components: a pure disc and a rounder halo.
The former dominates the true disc in regions of small accelerations, where
it controls the z-dynamics in the disc (disc flaring etc.);
it has a finite total mass. It also dominates the round component near the
 centre where the geometry is nearly planar. The second
 component controls motions far from the 
plane, has a total enclosed mass that diverges linearly with radius, and
determines the rotation curve at large radii. Its ellipticity may be 
appreciable at small radii but vanishes asymptotically.
This prediction of MOND differs from what one expects from galaxy-formation
scenarios with dark matter. 
 Analyses to date, which, as they do, assume one component--usually with a constant 
ellipticity, perforce give conflicting results for the best-value
ellipticity, depending on whether they probe the disc or the sphere, small
radii or large ones.

\end{abstract}

\section {Introduction}
\label{introduction}
 The modified dynamics (MOND) predicts
 that the global structure of the Newtonist's
``dark halo'' in disc galaxies
 is rather more complex than is usually assumed for such halos.
It has a disc component and a rounder component with radius-dependent
flattening, becoming spherical at large radii.
This structure is at odds with what one naturally expects from
existing halo-formation scenarios under the dark-matter doctrine.
\par
The purpose of this paper is to point out the general properties of MOND 
``halos'' with the help of analytic results for simple models and 
situations.
\par
In section 2 I explain how fictitious dark matter arises in MOND and
describe its distribution for thin-disc galaxies. Exact expressions for the 
``dark-matter'' distribution in  Kuzmin discs, for which
the MOND problem can be
solved exactly, are given in 
section 3. In section 4 I discuss generalizations of these
 results; in particular, the expected $z$-distribution of the 
``dark matter'' within the disc. Section 5 touches briefly on implications
for halo-shape analyses.
\section {Fictitious dark matter according to MOND}
\label{dm}
\par
I work with the formulation of MOND as a modification of Newtonian gravity
given in \cite{bm}
whereby the gravitational potential, $\psi$, of a density distribution 
$\rhos$ satisfies
\beq \div[\mu(\ags/\a0)\grad\psi]=4\pi G\rhos \label{Ii} \eeq
instead of the Poisson equation;
here, $\a0$ is the acceleration constant of MOND,
 $\mu(x)\approx x$ for $x\ll 1$, and $\mu(x)\approx 1$ for $x\gg 1$.
A Newtonist, who assumes that $\psi$ solves the Poisson equation, will 
deduce a mass density $\rho=(4\pi G)^{-1}\Delta\psi$, and will thus 
find that he requires ``dark matter'' with density
\beq \rho\_{D}=(4\pi G)^{-1}\Delta\psi-\rhos. \label{kapiyu} \eeq 
\par
Here I investigate the distribution of this fictitious mass density for
disc galaxies. For the sake of concreteness I consider a thin, axisymmetric, 
pure-disc model. The generalization to a galaxy model containing a bulge is 
straightforward and, in any case, does not  qualitatively change the results. 
\par
From eqs. (\ref{Ii}) and (\ref{kapiyu}) with $\rhos=0$,
the ``dark matter'' has a component outside the disc
 \beq \rho\_{D}=-(4\pi G)^{-1}\grad\ell n~\mu(\ags/\a0)\cdot\grad\psi
\label{buoteq} \eeq 
It is asymptotically spherical with 
$\rho\_{D}\rightarrow (4\pi)^{-1}(M\a0/G)^{1/2}R^{-2}$ at large radii $R$;
 $M$ is the total mass of the disc.
 This is the component that
gives the asymptotically flat rotation curve. There is
 also a thin-disc
component of ``dark matter'' because the jump condition on the potential is
different from the Newtonian jump for the same surface density.
In MOND, we have across
a thin layer of true surface density $\Sigmas$
\beq \d[\mu(\grm/\a0)\grm_z]=4\pi G\Sigmas, \label{gytda} \eeq 
where $\vg=-\grad\psi$ is the (MOND) acceleration, $\grm_z$ its
component perpendicular to the disc, and $\d[]$ is the jump from one side
of the disc to the other. 
A Newtonist who measure the the acceleration field
 will deduce a surface density 
$\Sigma= (4\pi G)^{-1}\d[\grm_z]$. For the $\pm z$-symmetric case,
 which I assume all along, we have
\beq \Sigma=\Sigmas/\mu(\grm^{\pm}/\a0), \label{laio} \eeq
or a ``dark-mater'' surface density
\beq \Sigma\_{D}(r)=[1/\mu(\grm^{\pm}/\a0)-1]\Sigmas\equiv\eta(r)\Sigmas(r),
 \label{lalu} \eeq
where $\grm^{\pm}$ is the total (MOND) acceleration just outside the disc.
 It follows
 from the basic assumptions of MOND that
 $\eta>0$, that it is very small where $\grm\gg \a0$, and that 
asymptotically $\eta\propto r$.
 We also see, generally, that  $\Sigma\_{D}$ is non zero only where there 
is a true disc. To get
 $\eta(r)$ we first have to solve the MOND equation for $\vg$.
\par
Note that for a one-dimensional 
planar mass layer all the ``dark matter'' lies in the layer
itself; outside, the MOND acceleration is constant and so gives
$\rho\_{D}=0$. This means that near the centre of the disc, where the geometry is
nearly planar, most of the dark matter lies in the disc and the equipotentials
are highly flattened.
The Newtonist's central surface density of the disc is 
\beq \Sigma_0= \i(\tilde y)\Sigmas_0>\Sigmas_0, \label{gurtewa} \eeq
where $\tilde y=2\pi G\Sigmas_0/\a0$, $\Sigmas_0$ is the true
central surface density; the function $\i(y)$ is defined by
 $\i(y)\equiv I^{-1}(y)/y$,
where  $y=I(x)\equiv x\mu(x)$ (so we have $\i(y)=1/\mu[x(y)]$). 
The argument of $\nu(y)$ is
always of the form $\gnrm/\a0$, where $\gnrm$ is some Newtonian
acceleration, while $x$ is of the form $\grm/\a0$ where $\grm$ is a MOND
acceleration.
 Regardless of the exact form of $\mu(x)$ we have
$\i(y)\approx 1$ for $y\gg 1$, and $\i(y)\approx y^{-1/2}$ for $y\ll 1$.
\par
For the widely used form of $\mu(x)=x(1+x^2)^{-1/2}$ we have
\beq \i(y)=[1/2+(y\^{-2}+1/4)\^{1/2}]\^{1/2}. \label{iota} \eeq
Equation(\ref{gurtewa}) is an example where the dark-matter factor
 $1/\mu(x)$, as appears e.g. in eqs.(\ref{buoteq}) and (\ref{laio}), 
calculated at some MOND field value $x=\grm/\a0$, can be expressed in terms of
the Newtonian-field variable $\gnrm/\a0$. This is possible for all
one-dimensional configurations where an exact algebraic relation exists
between the Newtonian and MOND accelerations
 $\vg\_{{\rm N}}=-\grad\f$ and
 $\vg=-\grad\psi$ respectively: 
\beq\mu(\grm/\a0)\vg=\vg\_{{\rm N}},  \label{IIi}\eeq
or inverting,  the MOND acceleration is given by
\beq\vg=\i(\gnrm/\a0)\vg\_{{\rm N}}.\label{acceler}\eeq
\par
I now demonstrate all the above with analytic expressions for Kuzmin discs.
\section {Analysis of the Kuzmin disc}
\label{kuzmin}
Kuzmin discs have a
Newtonian gravitational potential
\beq\f=-MG/[r\^{2}+(\abs{z}+h)\^{2}]\^{1/2}, \label{IIIii}\eeq
where $M$ and $h$ are the mass and scale length of the disc
(we use cylindrical coordinates $r,z,\theta$).
The potential above the disc ($z>0$) is that of a point mass $M$
 at $-\vh \equiv(0,-h,0)$;
the potential below the disc is produced by the same mass
at $\vh$. The surface density,
 $\Skr$, matches the jump in the $z$-gradient
of the potential
\beq\Skr=(2\pi G)\^{-1}\pd{\f}{z}\biggr\vert\_{z=0\^{+}}=
Mh/2\pi\rhsq\^{3/2}=\Sigma_0(1+u^2)^{-3/2},    \label{IIIiii}\eeq
where $\Sigma_0=M/2\pi h^2$, and $u=r/h$.
Everywhere outside the disc the equipotential surfaces are
concentric spheres centred at $\pm\vh$. Because of this one readily
gets the exact MOND solution (\cite{brada}). 
There the algebraic relations (\ref{IIi})(\ref{acceler}) hold
exactly, where here we now have
\beq\vg\_{{\rm N}}(\vR)=-MG(\vR\pm\vh)/\abs{\vR\pm\vh}\^{3}
 \label{IIIv}\eeq
for the Newtonian acceleration field above (+) and below ($-$) the disc.
We use the parameter  $\z\equiv MG/h\^{2}\a0$ to
 measures how deep in the MOND regime 
the disc is. For $\z\ll 1$ the whole disc is in the MOND regime, and we can
write everywhere outside the disc
\beq\psi \approx(MG\a0)\^{1/2}ln[r\^{2}+(\abs{z}+h)\^{2}]\^{1/2}.
 \label{IIIvi}\eeq
Otherwise this expression is only valid asymptotically (for $R/h\gg \z^{1/2}$).
\par
The Newtonist's mass distribution comprises a 
 disc whose surface density is
\beq  \Sigma(r)=\i(y)\Skr,  \label{sigma}\eeq
where 
\beq y(r)\equiv \gnrm(r,0\^{+})/\a0=MG/\a0\rhsq=\z/(1+u^2).
\label{kalapo} \eeq
So for Kuzmin discs
\beq \eta(r)=\{\i[\z/(1+u^2)]-1\}. \label{sigd} \eeq
For large values of $y$ (the Newtonian regime),
 as are found in the inner parts of a disc with 
$\z\gg 1$, we have 
$\eta\ll 1$. When $\z\ll 1$, or at large radii for all $\z$,
we have
\beq \eta(r)\approx \sqrt{{1+u^2\over \z}}-1\gg 1, \label{buyio} \eeq
and the ``dark disc'' dominates. (For the Kuzmin disc where
$\Skr\propto r^{-3}$ at large $r$, $\Sd\propto r^{-2}$ there, and the 
``dark disc'' mass diverges logarithmically.)
For the choice $\mu(x)=x(1+x^2)^{-1/2}$ we have
from eq.(\ref{iota})
$\eta=[1/2+(y\^{-2}+1/4)\^{1/2}]\^{1/2}-1$.

\par
In addition we need outside the disc a rounder component of
 "dark-matter" with density
\beq \rho\_{D}(r,z)=-(4\pi G)^{-1}\div\vg=-{M\over 2\pi h^3}\z^{-3/2}
\hat y^{5/2}\i'(\hat y) , 
 \label{poltuta} \eeq
where $\hat y=\gnrm/\a0=\z[u^2+(1+v)^2]^{-1}$, $v=\abs{z}/h$, and
$\i'(y)=d\i(y)/dy$. We have
$\rho\_{D}\propto R^{-2}$ asymptotically.
The equi-density surfaces of this distribution coincide with the 
equipotentials of $\psi$ and $\f$: they are spheres centred
at $\pm\vh$: all quantities are functions of $u^2+(1+v)^2$.
As expected the eccentricities are large near the centre ($u,v\lesssim 1$). 

\section {Generalizations}
\label{generalizations}
\par
For a thin disc with an exponential surface density profile the algebraic 
relation between Newtonian and MOND
 accelerations, eqs.(\ref{IIi}) and (\ref{acceler}), is a very good 
approximation (as shown numerically in \cite{brada}).  
Expression (\ref{sigd}) for $\eta$ is thus still a good approximation
with $ y(r)= \gnrm(r,0\^{+})/\a0$; but  $\gnrm$ is now the 
Newtonian acceleration for the exponential disc. Using the analytic 
expression for $\grm_r$ from \cite{freeman}, I find
$y(r)=\z[exp(-2s)+q^2(s/2)]^{1/2}$,
with $s=r/h$, $q(x)=x[I_0(x)K_0(x)-I_1(x)K_1(x)]$, and $I_i,~K_i$ 
modified Bessel functions.
\par
I have so far discussed only the z-integrated ``dark-matter'' surface
density. We need to know how
it is distributed in z to predict dynamics in the disc.
In the limit where the disc is very thin
the term with the radial part of the divergence in eq.(\ref{Ii})
 can be neglected in the bulk of the disc (where $d\grm_z/dz\gg d\grm_r/dr$),
 and we write
\beq \rhos\approx(4\pi G)^{-1}{\partial\over\partial z}[\mu(\grm/\a0)\grm_z].
\label{mitauta} \eeq
Or, we can turn this into an expression for the integrated surface
density between $\pm z$: $\Sigmas(r,z)\approx (2\pi G)^{-1}\mu(\grm/\a0)\grm_z$;
$\grm=(\grm_r^2+\grm_z^2)^{1/2}$.
In our approximation $\grm_r$ is independent of $z$,
but $\grm_z$ varies strongly with $z$.
The Newtonist's integrated surface density is
 $\Sigma(r,z)\approx (2\pi G)^{-1}\grm_z=\Sigmas(r,z)/\mu(\grm/\a0)$.
Just outside the disc $\grm_z$ becomes independent of $z$ and we get the 
saturated ratio of Newtonist-to-true surface density as in eq.(\ref{laio}).
If we are in a regime where $\grm_r\gg\grm_z$ for all $z$ (this is generally
valid beyond a few scale lengths) than $\grm$, and hence $\mu$,
 is approximately $z$ independent,
and the ``dark matter'' $z$ distribution is proportional to that of
 the true density.
Otherwise the ``dark matter'' is more concentrated near the mid-plane than
 the true density, since $\grm_z$, and hence $\mu$, increases with $z$.

\section {Discussion}
\label{discussion}
\par
In recent papers \cite{sack}, \cite{om}, and \cite{ibata} note that
 different techniques for estimating the ellipticity of galactic dark halos
 give systematically different results. These analyses were applied to 
different galaxies, so, strictly speaking, are not in conflict with each
 other. It is, however, alarming that the degree of flattening deduced is so
strongly correlated with the type of technique used.  
\par
But, if MOND is basically correct one cannot
successfully model the dark-matter distribution in disc galaxies by a 
single-component, elliptical halo of constant ellipticity.
Probing the 
galaxy far from the plane and from the centre will result in rounder halos
as in \cite{ibata}.
Probing near the centre might give appreciable
 eccentricities. For example, the considerable ellipticity that \cite{sackett}
 find for the halo of NGC 4650A reflects mostly constraints from fits to
 the rotation curve of the smaller, main disc. NGC 4650A is a 
low-surface-density galaxy, $\z\ll 1$, requiring much
dark matter even at small radii. This, we saw, is concentrated in the disc,
and so attempts to put it in a halo around the central disc requires a highly
 flattened halo.
Indeed, \cite{ms} who analyzed the rotation curves of the main disc and the 
polar ring of this galaxy in a MOND potential derived from eq.(\ref{Ii}) 
find good agreement.

\par
Disc-flaring analyses might also indicate a 
highly flattened halo.
The main reason is that the MOND ``dark disc'' can greatly enhance the
self gravity of the disc and make
it dominant in determining the $z$ structure.
\par
It is impossible to make quantitative comparisons with published disc-flaring
analyses, as in \cite{bc} and \cite{olling}, without delving into
 the many details of the data and of the assumptions going into
 such analyses (e.g. the contribution of a stellar disc in addition to that
of the studied HI layer, vertical velocity dispersion, etc.).
As a demonstration, consider a pure disc of mass $M$,
scale length $h$, and surface density $\Sigma(r)=(M/2\pi h^2)f(r/h)$.
Take $r$ beyond a few scale lengths so that the accumulated mass is
roughly saturated. Also assume we are in the MOND regime ($\mu\ll 1$)
and at a radius on the flat part of the rotation curve. Neglecting the
confining effects of the inner disc and reckoning only with the 
local self gravity (i.e. assuming pure planar symmetry), the 
$z$ scale hight according to MOND is
\beq z_0\sim {\sigma_z^2(r)\mu(r)\over 2\pi G\Sigma(r)}, \label{kode} \eeq
 where $\sigma_z$ is the
 $z$ velocity dispersion.
 Applying instead the expression 
\cite{olling} uses to analyze HI flaring in NGC 4244,
which neglects disc self gravity and assumes that the confining agent
is a flattened halo, we have
$z_0\sim (\sigma_z/V)[2.4q/(1.4+q)]^{1/2}r$ (taking in his expression a core
 radius $r_c\ll r$), where $V$ is the (constant) rotational velocity at $r$,
and $q$ is the flattening. Equating the two expression we see that to 
get the scale hight for MOND with a halo one would need a $q$ value
satisfying $[2.4q/(1.4+q)]^{1/2}\sim(\sigma_z/V)(h/r)^2/f(r/h)$, where I made 
use of the MOND relation $V^2\approx MG/\mu r$.
So, for example, for an exponential disc, at 5 scale
 lengths, and with $\sigma_z/V=0.1$, a $q$ value of 0.24 is gotten.
\par
Using data for NGC 4244 from \cite{ollingdata} we have at the last observed
 point (13 kpc) a gas surface density of
 $\Sigma\sim 5\times 10^{-5} {\rm gr~cm^{-2}}$ (corrected for helium), 
the MOND factor $\mu\approx V^2/r\a0\sim 0.16$, and the planar velocity
 dispersion
$\sigma\sim 8 {\rm km~s^{-1}}$. Applying the MOND estimate for the scale 
hight eq.(\ref{kode}), and assuming that the $z$ dispersion
 equals the planar one,
gives $z_0\sim 1700$ pc compared with Olling's measured
 half width at half maximum
of 750 pc. We have to remember that $\sigma_z$ might be
somewhat different, that eq.(\ref{kode}) is only approximate, neglecting 
the global restoring force (see below), etc.
Without the MOND factor (``dark disc'' contribution) the 
expected value of $z_0$ is 6 times larger (i.e. $\sim 10$ kpc),
 so the true disc alone
is quite negligible in containing the HI layer in Newtonian dynamics.
 At 10 kpc the MOND value
of $z_0$ is $\sim 300$ pc ($\mu\sim 0.27$ is larger, and $\Sigma$
 is 10 times larger, than at 13 kpc) compared with a similar value 
in \cite{ollingdata}. So, the high degree of halo flattening that 
\cite{olling} finds for this galaxy may well be largely an artifact of the 
MOND ``dark disc''. 
\par
The above was only a demonstration of the importance of the
MOND effects in augmenting the self-gravity restoring force of the disc,
even in places where the surface density is small. In a detailed analysis
one must add the confining force, $g^*_z$, of the rounder
 ''halo'' (the global potential of the disc) to the local, self-gravity
component $g_z$. At very large radii, or beyond some sharp cut-off of the
disc, the former will dominate. We can estimate the relative importance
of the two effects through the parameter
$\beta\equiv g_z/g^*_z \approx 2\pi\Sigma(r)r^3/Mz_0$. This approximation is
good inasmuch as the Newtonian radial acceleration can be written as
$MG/r^2$, so the MOND radial acceleration is $MG/\mu r^2$, and so the
restoring force of a spherical potential would be  $zMG/\mu r^3$;  
also $g_z\approx 2\pi G\Sigma(r)/\mu(r)$.
Applying this to the above two radii in NGC 4244, with a MOND mass of
$3\times 10^9~M_\odot$, corresponding to an asymptotic velocity of 85 km 
s$^{-1}$, I get $\beta\approx 10$ at 10 kpc, and $\beta\approx 1.2$ at 13 kpc.
This means that self gravity dominates at 10 kpc and global forces can be
neglected, while at 13 kpc the two are of the same order, and the above MOND
estimate of $z_0$ at 13 kpc has to be decreased by a factor of about 2.


\end{document}